# Raman Spectroscopy of Diesel and Gasoline Engine-Out Soot Using Different Laser Power

Haiwen Ge[1,*], Zhipeng Ye[2], Rui He[2,*]

1. Department of Mechanical Engineering, Texas Tech University, Lubbock, Texas 79409, USA
2. Department of Electrical and Computer Engineering, Texas Tech University, Lubbock, Texas 79409, USA

**Abstract:** We studied engine-out soot samples collected from a heavy-duty direct-injection diesel engine and a port-fuel injection gasoline spark-ignition engine. The two types of soot samples were characterized using Raman spectroscopy with different laser power. A Matlab program using least-square-method with trust-region-reflective algorithm was developed for curve fitting. We used a DOE (design of experiments) method to avoid local convergence. This method was used for two-band fitting and three-band fitting. The fitting results were used to determine the intensity ratio of D and G Raman bands. We find that high laser power may cause oxidation of soot samples, which gives higher D/G intensity ratio. Diesel soot has consistently higher amorphous/graphitic carbon ratio and thus higher oxidation reactivity, in comparison to gasoline soot, which is revealed by the higher D/G intensity ratio in Raman spectra measured under the same laser power.

----------------------------------

* Corresponding author. E-mail: haiwen.ge@ttu.edu (Haiwen Ge); rui.he@ttu.edu (Rui He)

**Introduction**

Particulate matter (PM) suspended in Earth's atmosphere impacts climate change, health effects, ecological effects, and visibility. Soot from internal combustion engines contributes an important portion of PM. Stringent regulations have been applied to tailpipe soot emission, not only on soot mass but also on particle numbers. To meet the soot emission regulation becomes the bottle neck of modern engine development, especially for the diesel engines. Diesel particulate filter (DPF) has been extensively used in modern diesel engine, which traps many engine-out soot particles (Johnson and Joshi, 2017; Johnson, 2009). Gasoline particulate filter (GPF) (Chan et al., 2012; Fischer et al., 2017) is being considered as a very important solution for PM emission of gasoline direct-injection engines. To maintain the performance of DPFs and GPFs (such as pressure drop), periodic regeneration is necessary to remove the soot loading inside the filters. Thermal oxidation is a common active regeneration method to remove the soot accumulation. Another way to remove the soot loading inside the filters is through catalyzed reaction of soot with $NO_2$ in the exhaust gas, which is a passive way for regeneration. Under both circumstances, increasing oxidation reactivity of the soot particles is a critical step.

Soot particles are mainly composed of carbon (C). It is generated during combustion of hydrocarbon fuel through Hydrogen-Abstraction-Carbon-Addition (HACA) mechanism (Frenklach and Wang, 1991). The soot formation process includes nucleation, aggregation and surface growth. Soot oxidation by $O_2$, OH, and other species occurs simultaneously (Ge et al., 2008; Shi et al., 2011). The engine-out soot is the final particle survived through these soot formation and oxidation processes. Soot particles are usually assumed to have a spherical shape in simplified two-step models (Cantrell et al., 2009; Ge et al., 2009; Ge et al., 2010; Hiroyasu and Kadota, 1976; Kong et al., 2007; Shi et al., 2011), phenomenological multi-step models (Jia et al., 2009; Tao et al., 2006, 2007; Tao et al., 2009; Vishwanathan and Reitz, 2009, 2015), and comprehensive models (Frenklach and Wang, 1991; Kazakov and Frenklach, 1998; Kazakov et al., 1995; Kumar and Ramkrishna, 1996; Wen et al., 2005). Chainlike soot agglomerates of sizes up to a micrometer size range may exist. Carbon atoms at the edge sites of a soot particle are more reactive than those in the bulk. Amorphous soot particles have more edge sites than graphitic ones. More edge sites on a soot particle correspond to higher soot reactivity. Therefore, soot oxidation reactivity correlates with the ratio of amorphous to graphitic carbon. The presence of oxygen or volatile matter may also affect soot oxidation reactivity.

Raman spectroscopy is a nondestructive powerful tool for characterizing carbon materials and their nanostructures (Dresselhaus et al., 2005; Lui et al., 2014; Zhao et al., 2015). The Raman process is an inelastic light scattering process. Energy difference between the incoming and the scattered light corresponds to the energy of the molecular vibrations or elementary excitations, for instance, phonons or charge density waves. Raman spectroscopy is a sensitive technique for sample characterization. Changes in the atomic bonds or molecular structures of probed samples are usually reflected as a change in the frequency and/or linewidth of Raman peaks. Therefore, we can extract information about the structural order and/or the degree of graphitization of soot by Raman studies.

Carbon-related materials, for instance, carbon nanotubes and graphene show two major Raman bands, D and G bands, in the frequency range between 1000 and 2000 $cm^{-1}$. The D (for "Defect" or "Disorder") Raman band (centered at around 1350 $cm^{-1}$) is a disorder-induced band. Its intensity and linewidth usually increase with the density of defects/impurities in the sample. The G (for "Graphite") Raman band, centered at about 1600 $cm^{-1}$, is also sensitive to sample quality. It originates from the stretching of C-C bonds. In graphene layers the frequency and linewidth of the G band is an indicator of charge carrier density (Das et al., 2008; Yan et al., 2007) and strain (Huang et al., 2009) in the sample. The presence of the D band arises mainly from the non-$sp^2$ bonds at the edge sites of soot particles (Sadezky et al., 2005). These sites are more likely to interact with oxygen or other non-carbon elements in the air during soot oxidation processes (Sadezky et al., 2005). Therefore, the D band intensity can be linked to the density of carbon atoms at soot particle edges, and hence with the reactivity of soot.

In order to eliminate extrinsic effects such as laser power and laser focus, the ratio of D to G peak intensities, $I_D/I_G$, is used as the primary parameter for analysis. $I_D/I_G$ was found to correlate with aromatic layer size, $L_a$ (Ferrari and Robertson, 2000; Tuinstra and Koenig, 1970; Wei et al., 2015), with soot oxidation reactivity (Al-Qurashi and Boehman, 2008; Ivleva et al., 2007b; Knauer et al., 2009; Song et al., 2006; Song et al., 2007; Zhang and Boehman, 2013), with maturation of various coal samples and diesel engine soot aged at various temperatures (Fang and Lance, 2004), with fuel/oxygen ratio (Han et al., 2012), and with the temperature of treatment and different grade of carbon black (Pawlyta et al., 2015). Schmid et al. (Schmid et al., 2011) developed a multi-wavelength micro-Raman method to characterize soot structure and reactivity. The method is based on the dispersive character of the D band in Raman spectra, i.e., red shift at longer excitation laser wavelength. The method was applied to diesel soot analysis. It was found that the observed Raman difference integrals (difference between the integrated areal intensities of two Raman spectra) have a linear correlation with soot reactivity. Russo and Ciajolo (Russo and Ciajolo, 2015) used different excitation wavelengths in Raman spectroscopy to characterize the soot produced in premixed fuel-rich flames of different hydrocarbon fuels.

To determine the intensities of D band and G band as well as their ratio, curve fitting is required to reproduce the measured Raman spectra using several presumed distribution functions. After we fit the measured Raman spectra, we can extract the center frequency, bandwidth, and intensity of the observed Raman bands, whose exact values are not easy to determine by eyeballing. Information on the bandwidth and intensity ratio of D and G bands allows us to compare the oxidation activity of different types of soot. Sadezky et al. (Sadezky et al., 2005) conducted a comprehensive study of curve fitting for the Raman spectra of soot. They tested and compared nine different band combinations and demonstrated that the best fit to a wide range of soot samples was obtained with four Lorentzian (L) bands (G, D1, D2, and D4) and one Gaussian (G) band (D3). This fitting method has been extensively used for Raman spectra of soot (Al-Qurashi and Boehman, 2008; Atribak et al., 2010; Charbonneau and Wallace, 2009; Ess et al., 2016; Han et al., 2012; Ivleva et al., 2007a; Ivleva et al., 2007b; Knauer et al., 2009; Pawlyta et al., 2015; Soewono and Rogak, 2011). Soewono and Rogak (Soewono and Rogak, 2011) used a 2-band model and Sadezky (Sadezky et al., 2005) used a 5-band model, and they both concluded that soot from biodiesel has less structural order than the one from ULSD. Seong and Boehman (Seong and Boehman, 2013) compared different fittings (2L1G, 3L, 3L1G, 4L, 4L1G, 5L) and concluded that the 3L1G fitting is the best since its results indicate that soot oxidative reactivity is closely related to crystalline disorder. Based on 3L1G fitting, the crystalline width estimated using Knight and White's model matches fringe length given by TEM the best. Rusciano et al. (Rusciano et al., 2008) used the 5-band model to fit Raman spectra and found that it works well for soot but not for nano-sized organic carbon particles. Rusciano et al. found that five Gaussian (5G) gives the best fit. Herdman et al. (Herdman et al., 2011) considered 3- and 5-band models, as well as a 2-band Breit–Wigner–Fano (BWF) model. Russo and Ciajolo (Russo and Ciajolo, 2015) used different fitting methods and showed that BWF line shape provides an effective representation of the asymmetric broadening of the G peak. The mean $\chi^2$ values calculated at each excitation wavelength are 4.1 for 1BWF3L1G and 6L, 4.4 for 1BWF4L and 7.1 for 5L. Although 6L fitting method gives the lowest fitting error, the ratio of G to the peak at 1620 cm$^{-1}$ varies significantly, and thus the related Raman parameters show large uncertainties.

In the present work, Raman spectroscopy with different laser power was used to characterize the engine-out diesel soot and gasoline soot. Different from prior work that used 4-6 L or G peaks to fit Raman spectra, we used 2-3 peaks (L or G) in our fitting for simplification. We find that high laser power may cause oxidation of soot, which gives higher D/G intensity ratio. We also find that diesel soot has higher ratio of amorphous/graphitic C compared to gasoline soot, which is revealed by the higher D/G intensity ratio measured under the same laser power.

# 1 Materials and methods

## 1.1 Soot sample preparations

Soot particles were collected and placed on Teflon filters. The diesel soot was collected from a John-Deere diesel engine. It is a 2-valve 4.5L 4-cylinder heavy-duty direct-injection diesel engine for off-road applications (Ge and Cho 2018). Table 1 lists the key configurations and operating conditions of this engine. Standard diesel fuel was used. The Teflon filter was placed in the exhaust manifold of the diesel engine to collect soot samples.

The gasoline soot was collected from a SUV vehicle powered up by a 2.5L Nissan QR engine. It is an I4 DOHC port-fuel-injection spark-ignited engine. Its basic configuration is also listed in Table 1. Regular gasoline (87 octane) was used. The filter was placed in the tailpipe of the engine to trap the gasoline soot particles. During the soot particle collection procedure, FTP-75 (Federal Test Procedure) cycle was conducted on this SUV vehicle.

## 1.2 Raman measurements

Samples were measured at room temperature using a Horiba LabRam HR Raman Microscope system. We used 633 nm laser. By using a 100× objective lens, we were able to focus the laser beam to a spot with diameter less than 1 μm. The scattered light was dispersed by a 600-groove/mm grating (resolution of ~2 $cm^{-1}$) and detected by a thermoelectric cooled CCD detector. Different levels of laser power (0.01 mW, 0.1 mW, 1 mW) were applied to take Raman spectra.

## 1.3 Curve fitting method

Curve fitting was conducted using least-square-method with trust-region-reflective algorithm. Previous work found that curve fitting is sensitive to the initial guess of center, height, and width of different bands. This is because the present optimization problem is a multi-peak problem. Based on an arbitrary initial guess, the optimization may converge to local optima rather than a global optimum. To overcome this difficulty, the least-square-method is coupled with a Design of Experiments (DOE) method. Based on the pre-defined range of all fitting parameters, the DOE method generates a matrix that consists of multiple initial guess for the fitting parameters. The size of the DOE matrix is set to $N^2 \times N$, where $N$ is the total number of fitting parameters. The DOE matrix covers the whole domain of the $N$-dimensional space in a uniform way. Each vector of this DOE matrix is used as an initial guess for curve fitting. After $N^2$ curve fittings, a global optimum is obtained. Both two band fitting and three band fitting are considered. For three band fitting, Lorentzian line shape is used for G and D1 bands, and Gaussian line shape is used for D3 band. This fitting is denoted as "2L1G". Coefficient of determination (COD) is used to evaluate the effectiveness of the fittings.

## 2 Results and discussion

Figure 1 shows optical images of the two soot samples. The left one is from the diesel engine. The one on the right is from the gasoline engine. Figure 2 shows a comparison of normalized Raman spectra taken with different laser power levels from diesel engine soot. Background signal is subtracted from the raw data, and the spectra are then normalized to the G Raman band. After normalization, the Raman spectra measured at different power levels can be compared. The spectra are very typical for soot, with two bands in the first order region, the G band at around 1600 cm$^{-1}$ and the D band at around 1350 cm$^{-1}$. The spectra taken with 0.01 mW and 0.1 mW laser power are similar. However, both the D and G Raman bands in the spectrum taken at high laser power (1.0 mW) show a prominent redshift, indicative of lattice expansion under laser heating (Balandin, 2011) and/or tensile strain due to a change of C-C bonds (Huang et al., 2009; Mohiuddin et al., 2009). In addition, intensity of the D band under 1.0 mW laser power increases compared to those taken with 0.01 mW and 0.1 mW laser power, possibly due to higher degree of oxidation and disorder induced at high laser power. Quantitative analysis of $I_D/I_G$ is discussed below.

Figure 3 shows a comparison of normalized Raman spectra taken with different power levels from gasoline engine soot. It can be seen that the spectra taken with 1.0 mW and 0.1 mW laser power are similar. In comparison to the Raman spectra from diesel engine soot, spectra from gasoline engine soot differ in two ways. First, height intensities of the D and G bands are close to each other for gasoline engine soot (see Fig. 3), whereas the height intensity of the D band is higher than that of the G band for diesel engine soot (see Fig. 2). The higher D/G intensity ratio for diesel engine soot can be clearly seen in Fig. 4 which compares spectra from diesel and gasoline engine soot at the same laser power. This suggests that diesel engine soot may have higher degree of disorder (amorphous/graphitic carbon ratio) compared to the gasoline soot. Second, when laser power increases from 0.1 to 1 mW, the spectra from gasoline engine soot does not show much difference. However, in the spectra from diesel engine soot (Fig. 2), we can see that the D and G bands redshift and the D band intensity increases under high laser power of 1 mW. This result suggests that the diesel engine soot is probably more reactive in the air at high laser power which could give rise to an increase of D band intensity.

In the first step, two band fitting method (D1 and G bands) is considered. Three different functions, including Gaussian, Lorentzian, and Voigt, are used to fit the two bands. To identify the best fitting, different combinations of the three fitting functions (Gaussian, Lorentzian, Voigt) were tested on the spectrum taken with 0.1 mW from diesel soot sample. Effectiveness of the fitting curve is evaluated based on the value of COD($R^2$). "1" indicates a perfect fitting, whereas "0" indicates the worst fitting. The results are listed in Table 2. All fittings involving Gaussian lineshape show relatively low COD, while fittings involving Lorentzian lineshape consistently give the highest COD. Performance of Voigt function is between Gaussian and Lorentzian. The combination of Lorentzian+Lorentzian (2L) gives the best fit to the experimental data.

Figure 5 shows the fitting of diesel and gasoline soot spectra using two Lorentzian (2L). Both spectra were taken using 0.1 mW laser power. The fittings are reasonably good. Table 3 lists the curve-fitting results of all spectra from the two samples taken at different laser power using 2L fitting method. The $I_D/I_G$ of the diesel soot is higher than that of the gasoline soot. It implies that the oxidation reactivity of the diesel soot is higher than the gasoline soot. This may be related to the fact that the diesel engine combustion is dominated by diffusion flames and the soot is generated at richer condition than that from gasoline engine. Hence, the survived soot particles from the diesel engine usually have higher oxidation reactivity.

To further improve the curve-fitting, 3-band fitting method is considered by introducing a third band D3. We name this band D3 instead of D2 in order to be consistent with the assignment in the literature (Sadezky et al., 2005). Based on the two-band fitting above, D1 and G are best fit using Lorentzian line shape. We test the 3-band fitting by considering three functions (L, G, and V) for the D3 band. Table 4 shows the COD of fitting results of the spectrum from diesel soot measured using 0.1 mW laser power (the same spectrum we used to test the 2-band fitting above). We can see that Gaussian line shape is the best for fitting the D3 band as indicated by the highest COD. It implies that the D3 band is subject to a significant degree of inhomogeneous broadening. Lorentzian line shape is the worst for D3 fitting among these three functions. Voigt shows very similar performance to Gaussian for the D3 band. Because Gaussian is simpler than Voigt, two Lorentzian plus one Gaussian (2L1G) is used as the three-band fitting method. The COD value is improved from 0.9991421 for 2-band (2L) fitting to 0.9997781 for 3-band (2L1G) fitting.

Figure 6 shows 3-band fitting of Raman spectra from diesel and gasoline soot. We used the same spectra as we used for 2-band fitting shown in Fig. 5 for easy comparison. We can see that the 3-band fitting method is much better than the 2-band fitting. Table 5 lists the curve-fitting results of all spectra from both soot samples using 3-band fitting (2L1G) method. Compared to the spectra from diesel soot, spectra from gasoline soot exhibit larger widths of D1 band but smaller width of D3 band. Due to the presence of the D3 band, intensity of the G band is suppressed compared to the two-band fitting method, which leads to higher $I_D/I_G$ for the 3-band method (here $I_D$ refers to the total intensity of D1 and D3). Intensity ratio of D3 and G bands, which is denoted as $I_{D3}/I_G$, is listed in Table 5.

From both 2- and 3-band fitting results listed in Table 3 and Table 5, we can see that $I_D/I_G$ is lower for gasoline soot under the same laser power. It suggests that diesel soot has higher ratio of amorphous/graphitic C compared to gasoline soot. In addition, $I_D/I_G$ in general increases at higher laser power, suggesting that higher laser power (or higher temperature) could result in higher degree of soot oxidation reactivity.

## 3 Conclusions

We studied engine-out soot samples collected from a heavy-duty direct-injection diesel engine and port-fuel injection gasoline spark-ignition engine. The soot samples are characterized using Raman spectroscopy at different laser power. We find that high laser power may cause oxidation of soot, which gives higher D/G intensity ratio. Diesel soot has higher amorphous/graphitic C compared to gasoline soot, as revealed by higher D/G intensity ratio measured under the same laser power. The method can be applied to characterize different soot samples. The value of D/G intensity ratio may be linked to oxidation reactivity of the soot samples that determines the regeneration efficiency of the DPF/GPF. The measured D/G intensity ratio may also be used to develop advanced soot models.

It is worthwhile to note that fuel type, engine design, engine operating conditions including engine speed and engine load, and soot collection method may all influence soot composition and its resultant Raman spectra. Further studies are needed to investigate the impact of each parameter on soot formation.

**Acknowledgements**

The authors acknowledge Jeffrey S. Rademacher, Bryan T. Geisick, and Danan Dou of John Deere Power Systems for providing the diesel soot samples. Z. Y. and R. H. acknowledge the support by NSF (CAREER grant No. DMR-1760668).

**List of tables**

Table 1: Configurations of John Deere diesel engine and 2.5L Nissan QR engine

| **John Deere diesel engine** | | **2.5L Nissan QR engine** | |
|---|---|---|---|
| **Bore (mm)** | 106.5 | **Bore (mm)** | 97.0 |
| **Stroke (mm)** | 127.0 | **Stroke (mm)** | 45.0 |
| **Connecting rod (mm)** | 203.0 | **Connecting rod (mm)** | 143.05 |
| **Compression ratio** | 17:1 | **Compression ratio** | 9.5:1 |
| **Engine speed (rpm)** | 2400 | | |
| **Engine load** | Full load | | |

**Table 2**: Different two-band fitting methods for the diesel engine soot measured using 0.1 mW laser power.

| Band D1 | Band G | COD($R^2$) |
|---|---|---|
| Gaussian | Gaussian | 0.9976601 |
| Gaussian | Lorentzian | 0.9978074 |
| Gaussian | Voigt | 0.9974512 |
| Lorentzian | Gaussian | 0.9988361 |
| Lorentzian | Lorentzian | 0.9991421 |
| Lorentzian | Voigt | 0.9987685 |
| Voigt | Gaussian | 0.9981657 |
| Voigt | Lorentzian | 0.999068 |
| Voigt | Voigt | 0.9986337 |

**Table 3**: Curve-fitting results of all spectra from both soot samples using two-band (2L) fitting method.

|  | DIESEL 0.01 MW | DIESEL 0.1 MW | DIESEL 1.0 MW | GASOLINE 0.1 MW | GASOLINE 1.0 MW |
|---|---|---|---|---|---|
| **D1 CENTER** | 1343.0 | 1338.9 | 1331.6 | 1342.8 | 1343.2 |
| **D1 WIDTH** | 103.2 | 100.8 | 106.4 | 117.4 | 111.0 |
| **G CENTER** | 1599.4 | 1597.8 | 1583.4 | 1592.6 | 1593.5 |
| **G WIDTH** | 42.7 | 43.7 | 45.6 | 43.3 | 41.5 |
| $I_D/I_G$ | 2.555 | 2.510 | 2.606 | 2.357 | 2.426 |
| **COD($R^2$)** | 0.9984592 | 0.9988198 | 0.9991723 | 0.9977284 | 0.9983489 |

**Table 4**: Comparison of different three-band fitting methods for spectrum from the diesel soot measured using 0.1 mW laser power.

| Band D1 | Band G | Band D3 | COD($R^2$) |
|---|---|---|---|
| Lorentzian | Lorentzian | Gaussian | 0.9997781 |
| Lorentzian | Lorentzian | Lorentzian | 0.9996907 |
| Lorentzian | Lorentzian | Voigt | 0.9997778 |

**Table 5**: Curve-fitting results of all spectra from both soot samples using 3-band fitting (2L1G) method.

|  | Diesel 0.01 mW | Diesel 0.1 mW | Diesel 1.0 mW | Gasoline 0.1 mW | Gasoline 1.0 mW |
|---|---|---|---|---|---|
| **D1 center** | 1340.8 | 1336.5 | 1329.7 | 1341.9 | 1342.3 |
| **D1 width** | 93.6 | 91.2 | 97.9 | 106.5 | 99.9 |
| **D3 center** | 1548.4 | 1544.6 | 1540.2 | 1553.5 | 1554.5 |
| **D3 width** | 58.0 | 59.5 | 60.2 | 55.2 | 52.3 |
| **G center** | 1607.9 | 1606.5 | 1591.8 | 1599.0 | 1601.6 |
| **G width** | 30.8 | 31.0 | 33.9 | 34.4 | 29.6 |
| $I_D/I_G$ | 4.222 | 4.301 | 4.630 | 3.592 | 4.257 |
| $I_{D3}/I_G$ | 0.498 | 0.527 | 0.544 | 0.377 | 0.527 |
| **COD($R^2$)** | 0.999335 | 0.9997781 | 0.9997782 | 0.99826228 | 0.9994198 |

**List of figures**

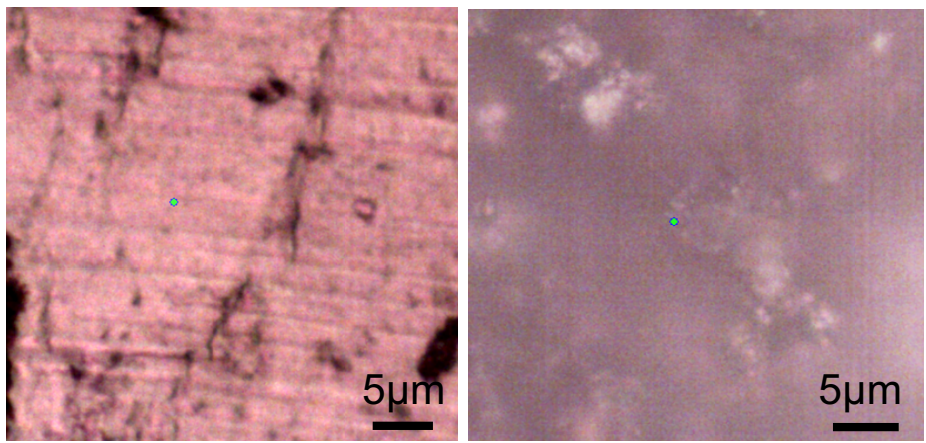

**Fig. 1**: Raw image of the diesel soot sample (left) and gasoline soot sample (right). The green dot shows the spot where we focused our laser on.

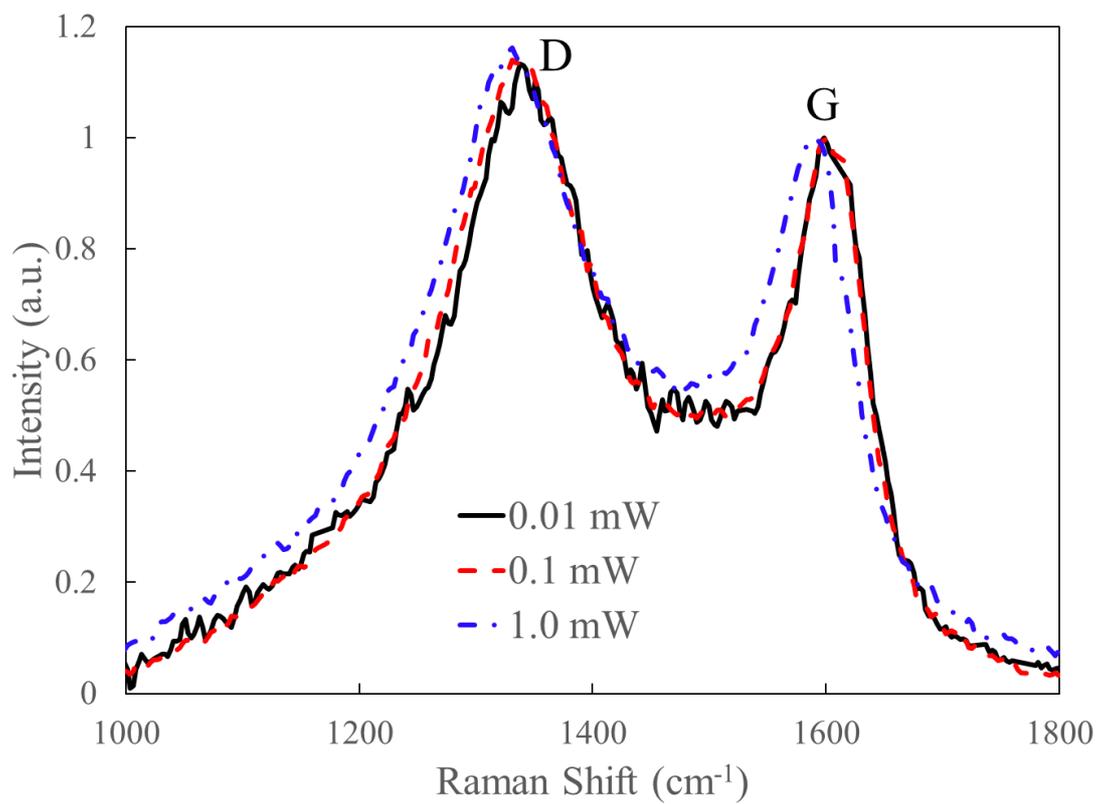

**Fig. 2**: Raman spectra from diesel engine soot taken with different laser power.

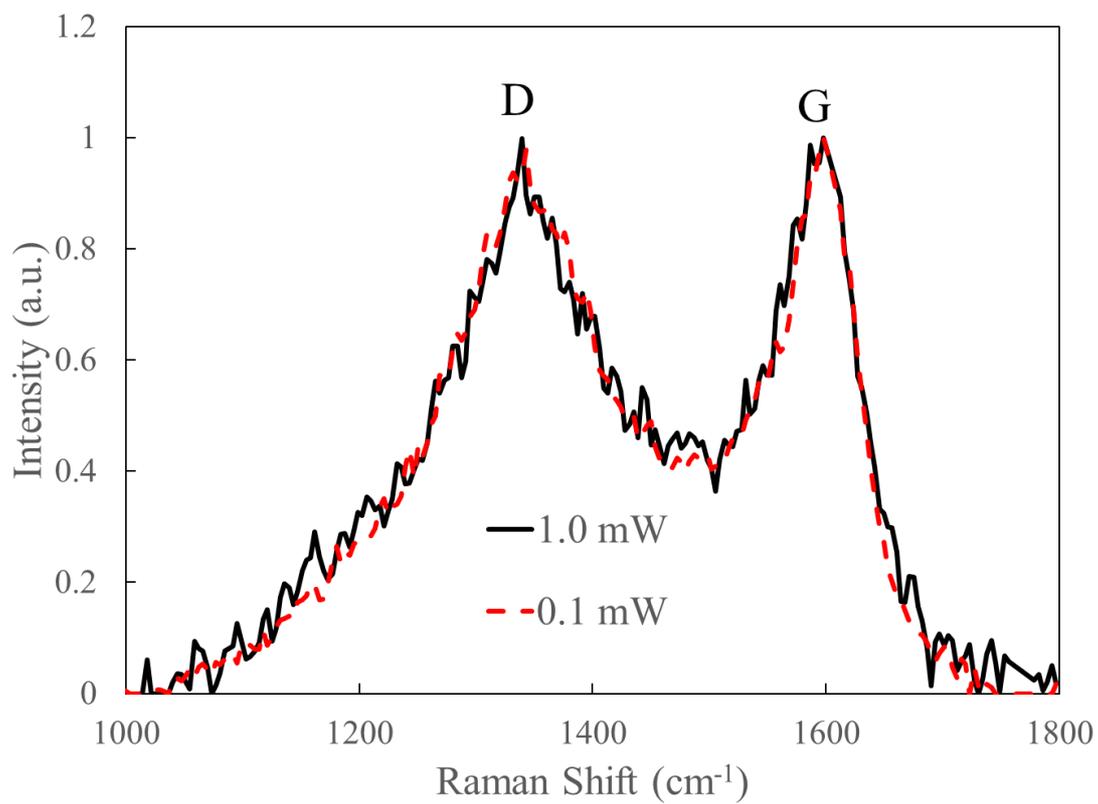

**Fig. 3**: Raman spectra from gasoline engine soot taken with different laser power.

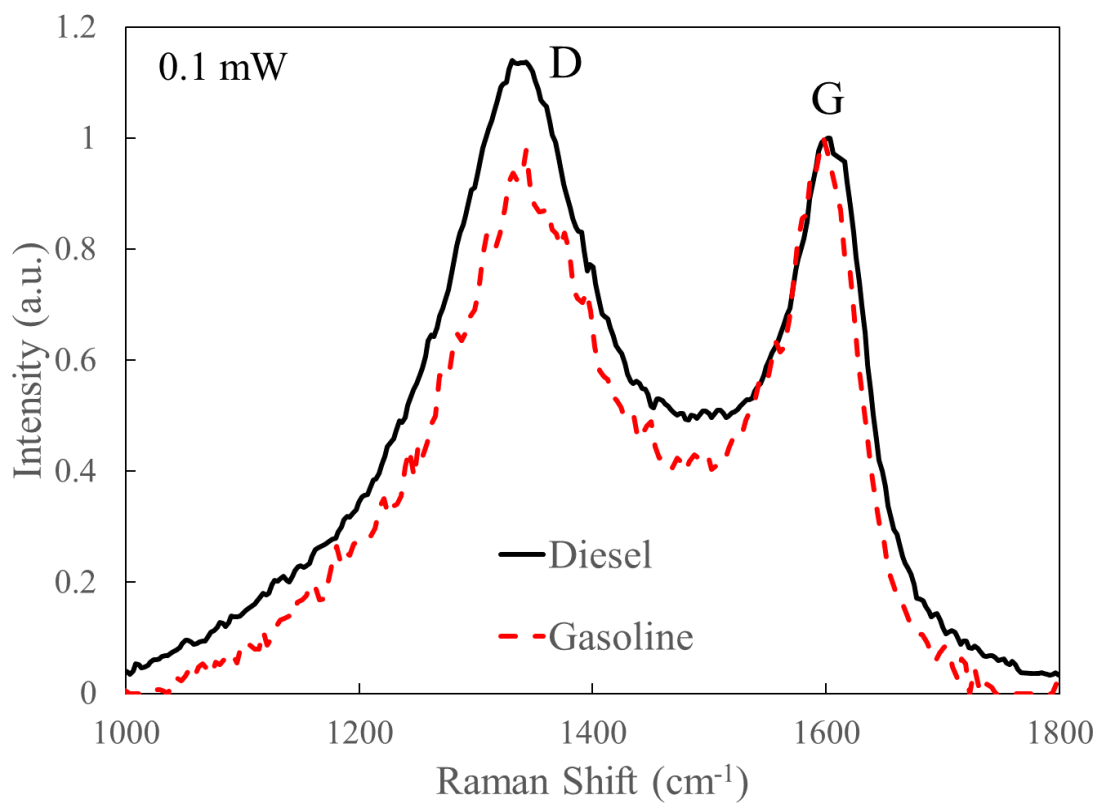

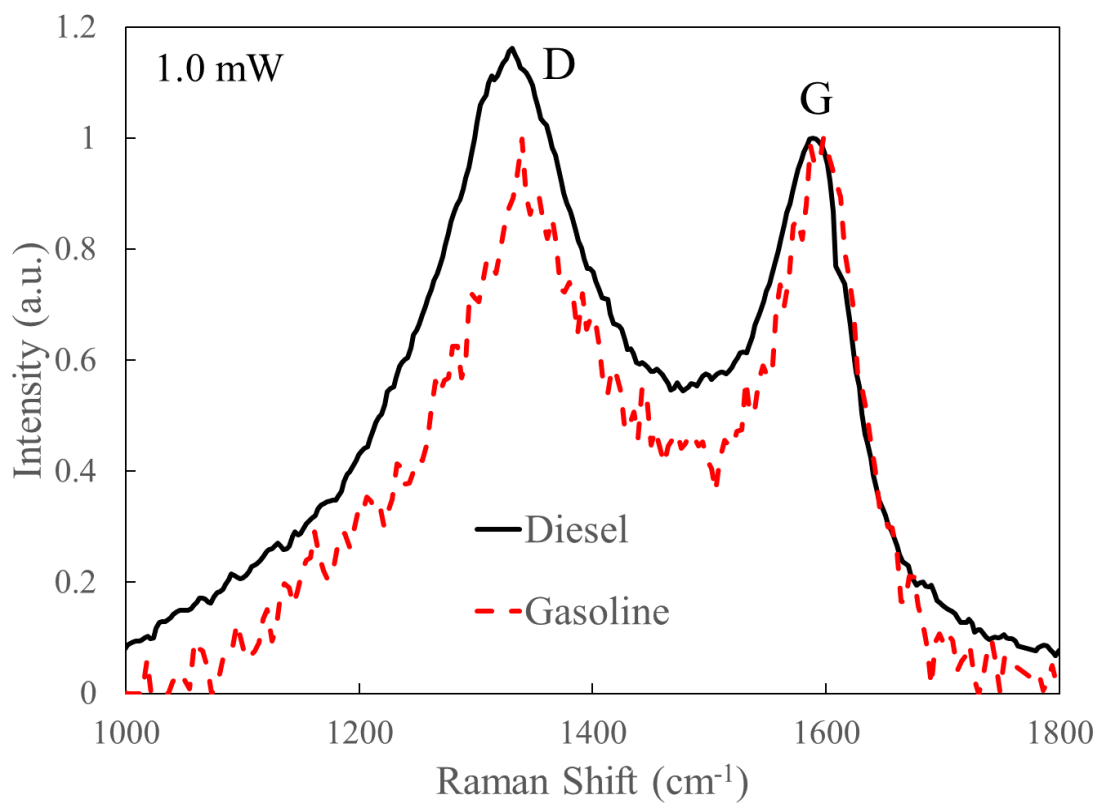

**Fig. 4**: Raman spectra of diesel and gasoline engine soot taken at laser power of 0.1 mW (left) and 1.0 mW (right).

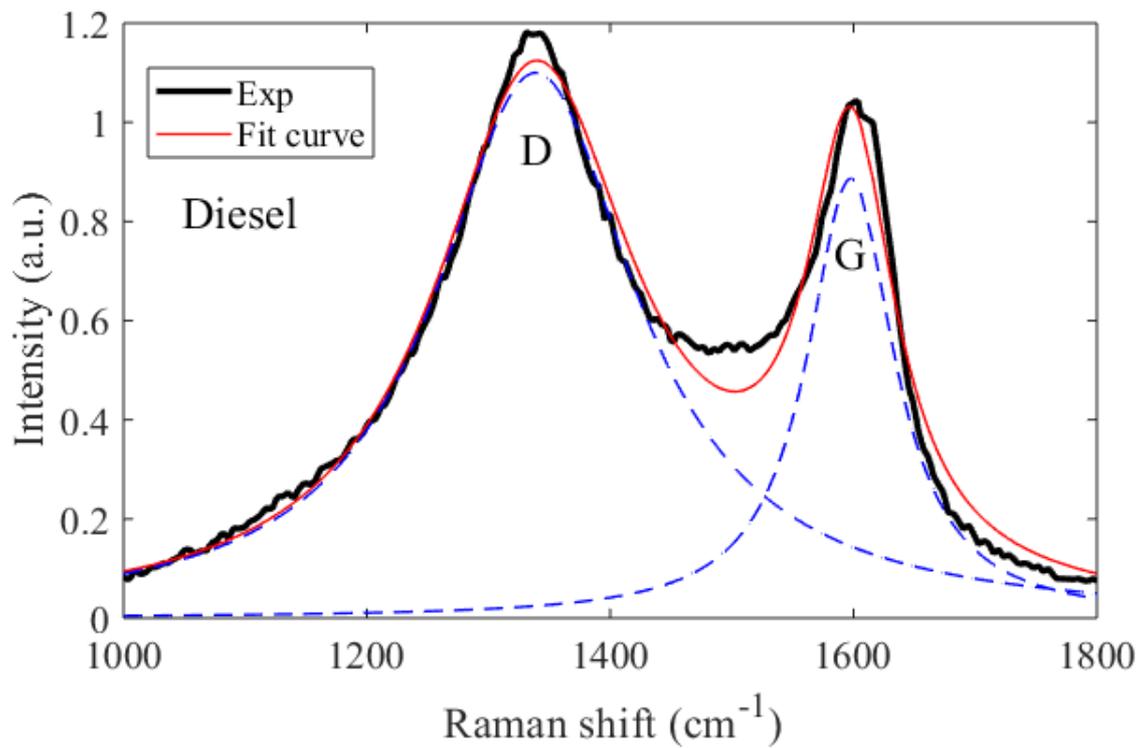

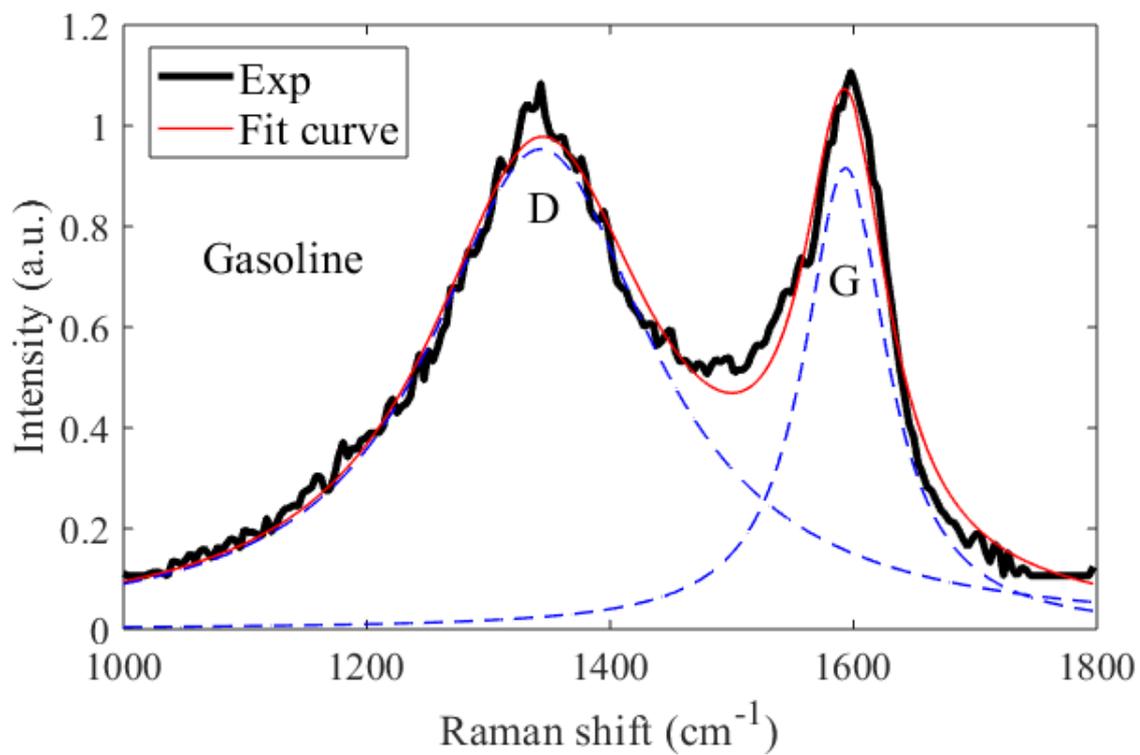

**Fig. 5**: Two-band fitting of Raman spectra from diesel and gasoline soot taken with 0.1 mW laser power using two Lorentzian.

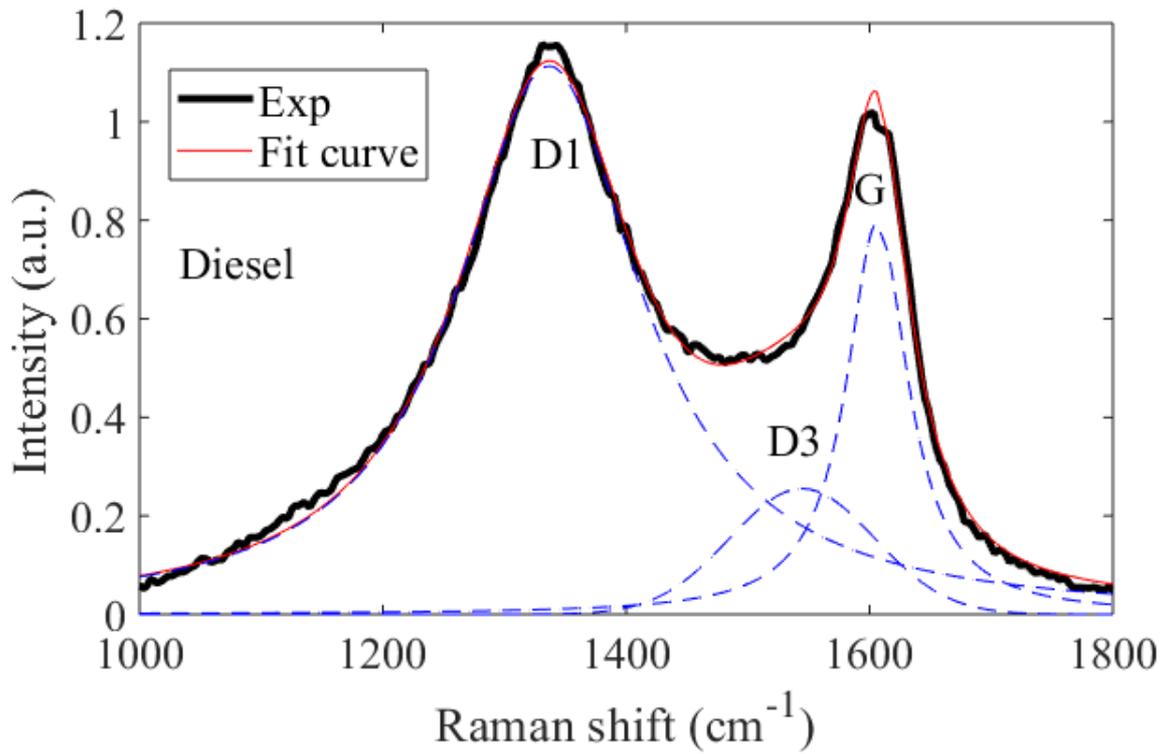

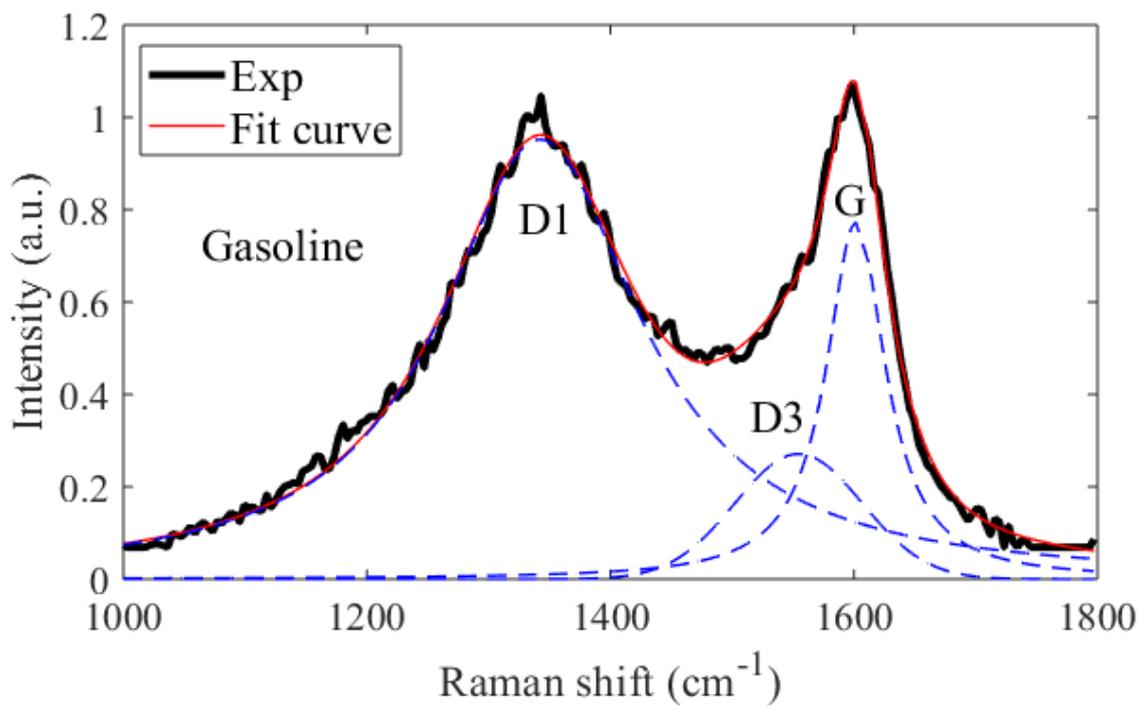

**Fig. 6**: Three-band fitting of spectra from diesel and gasoline soot taken with 0.1 mW laser power using two Lorentzian (D1 and G) and one Gaussian (D3).